\documentclass[doublecol, figures]{epl2}
\usepackage{amsmath, amssymb, mathrsfs}

\title{Tailoring the thermal Casimir force with graphene}

\author{Vitaly Svetovoy \inst{1} \and Zakaria Moktadir \inst{2}
\and Miko Elwenspoek \inst{1,3} \and Hiroshi Mizuta \inst{2,4}}

\shortauthor{Vitaly Svetovoy \etal}

\institute{\inst{1} MESA$^+$ Institute for Nanotechnology, University of
Twente, PO 217, 7500 AE Enschede, The Netherlands\\
\inst{2} University of Southampton, Highfield, Southampton, SO17
1BJ, United Kingdom\\
\inst{3} FRIAS, University of Freiburg, 79104 Freiburg, Germany\\
\inst{4} School of Materials Science, Advanced Institute of Science
and Technology (JAIST), Japan}




\abstract{The Casimir interaction is omnipresent source of forces at
small separations between bodies, which is difficult to change by
varying external conditions. Here we show that graphene interacting
with a metal can have the best known force contrast to the
temperature and the Fermi level variations. In the distance range
$50-300$ nm the force is measurable and can vary a few times for
graphene with a bandgap much larger than the temperature. In this
distance range the main part of the force is due to the thermal
fluctuations. We discuss also graphene on a dielectric membrane as a
technologically robust configuration.}

\begin{document}

\maketitle

\section{Introduction}

The Casimir force \cite{Cas48} manifests itself at short distances
($<1$ $\mu$m) as a result of the electromagnetic interaction between
neutral bodies without permanent polarizations. For two ideally
reflecting parallel plates separated by a distance $a$, this force is
given by: $F_C=(\pi^2/240)(\hbar c/a^4)$. The universal character of
the force stimulated active development of the field \cite{Cap07}
with applications in physics, biology, and technology.

The Lifshitz theory \cite{Lif55} gives the most detailed description
of the force. According to this theory, current fluctuations (quantum
and classical) in the bodies are responsible for the force.
Fluctuations in a wide range of frequencies give significant
contribution to the force. For this reason it is difficult to change
the force at will as one has to modify the dielectric response of
interacting materials in a wide range of frequencies.
Hydrogen-switchable mirrors did not show observable contrast to the
Casimir force \cite{Ian04}. It was demonstrated that the force
between indium tin oxide (ITO) and a gold surface is 50\% smaller
than it is between two Au surfaces \cite{Man09}. For the same
material the best result was found for the phase-changing material
(Ag-In-Sb-Te) with 20\% difference between amorphous and crystalline
phases \cite{Tor10}. {\it In situ} modulation of the force between a
gold sphere and a silicon membrane \cite{Che07} was shown to 1\%
level when the carrier density was changed optically by 4 orders of
magnitude.

The force measured in modern experiments is mainly the result of
quantum fluctuations whilst the force due to classical
fluctuations (thermal Casimir or Lifshitz force) was measured only
recently between an ultracold atomic cloud and a sapphire
substrate \cite{Obr07}, and between two Au surfaces \cite{Sus11}.
The thermal fluctuations dominate the force at large distances
$a\gtrsim \hbar c/T$ ($k_B=1$) where the force itself is extremely
weak and approaches the Lifshitz limit \cite{Lif55}. Between two
metals this limit is given by:
\begin{equation}\label{Lif}
    F_{L}=\frac{T\zeta(3)}{8\pi a^3}, \ \ \ a\gg \lambda_T=
    \frac{\hbar c}{T},
\end{equation}
where $\lambda_T$ is the thermal wavelength and $\zeta(x)$ is the
zeta-function.

In this paper we show that significant variation (up to 5 times) of
the total Casimir force is possible for graphene with a bandgap
$2\Delta\gg T$. The force changes in response to the variation of the
Fermi level mainly due to the change of its thermal part. It can be
realized at the distance range $a=50-300$ nm, where the force is well
measurable.

Graphene, a single layer material with carbon atoms arranged in a
honeycomb lattice, attracted enormous attention \cite{Nov04,Cas09}.
Unusual electronic properties of graphene are due to massless
relativistic dispersion of electrons at low energies
\cite{Cas09,Kot10}. The Casimir/van der Waals interaction of
graphene was mainly discussed at zero temperature
\cite{Bor06,Bor09,Ser10} with the conclusion that the force due to
graphene is weak in comparison with the interaction of bulk bodies.

An important development was made by G\'{o}mez-Santos \cite{Gom09} at
finite temperature. It was argued that at $T=0$ graphene is a
critical system, with no characteristic length scale. At nonzero $T$
this scale is given by the thermal length $\xi_T=\hbar v_F/T$, where
$v_F\approx 10^6$ m/s is the Fermi velocity in graphene. It was found
that in the long distance limit the force between two graphene sheets
is given by the same Eq. (\ref{Lif}) but this equation is true for
much shorter distances $a\gg \xi_T$. At room temperature the scales
$\xi_T$ and $\lambda_T$ are 25 nm and 7.6 $\mu$m respectively. This
property makes the thermal Casimir force operative for separations in
the $50 - 300$ nm range, which are readily accessible using an atomic
force microscope (AFM) or other force measuring techniques (see
\cite{Cap07} for a review).

Graphene is a promising material for the development of
high-performance electronic devices \cite{Bol08} but pristine
graphene is a semimetal with zero bandgap \cite{Cas09}. The major
challenge of graphene electronics is to open an energy bandgap
\cite{Han07}. As we will see later, the bandgap is also important
for tailoring the Casimir force by electronic means. Significant
progress has been made in this direction. For instance, epitaxially
grown graphene on SiC has a gap of $2\Delta\approx 0.26$ eV
\cite{Zho07}. Opening a bandgap was also demonstrated by water
adsorption \cite{Yav10} and patterned hydrogen adsorption
\cite{Bal10}. Graphene nanomesh proved to generate bandgaps with
values depending on the mesh density \cite{Par08,Bai10,Kim10}. Very
recently, an efficient way to fabricate graphen nanomesh was
developed \cite{Zha11}. In the present work we will assume the
presence of a gap without specifying its origin.

\section{Graphene on a substrate}

We consider here the interaction between two plates 1 and 2 having
dielectric functions $\varepsilon_1(\omega)$ and
$\varepsilon_2(\omega)$, respectively. In contrast with \cite{Gom09}
graphene is not free standing but covers the plate 1. As we will see
it has significant influence on the system. The case of suspended
graphene is reproduced by taking $\varepsilon_1(\omega)=1$. The
Lifshitz formula \cite{Lif55} expresses the force between two
parallel plates via their reflection coefficients. If graphene sheet
has the two dimensional (2D) dynamical conductivity $\sigma$, then
the reflection coefficient of the plate with graphene (for p
polarization) is given by \cite{Fal07,Sta08}:
\begin{equation}\label{rp}
    r_{1}=\frac{k_0\varepsilon_1-k_1+\left(4\pi\sigma/\omega\right)k_0
    k_1}{k_0\varepsilon_1+k_1+\left(4\pi\sigma/\omega\right)k_0
    k_1}.
\end{equation}
Here the normal components of the wave vectors in vacuum and in the
substrate are $k_0=\sqrt{\omega^2/c^2-q^2}$ and
$k_1=\sqrt{\varepsilon_1\omega^2/c^2-q^2}$, respectively, where
$\textbf{q}$ is the wave vector along the plate. In the $T=0$ limit
the graphene conductivity is $\sigma\sim e^2/\hbar$ for frequencies
up to near UV \cite{Fal07,Nai08,Daw08}. It means that the reflection
coefficient gets only a small correction $\sim\alpha=e^2/\hbar
c=1/137$ due to the presence of graphene on the dielectric
substrate. This explains a weak force between two graphene sheets
\cite{Bor06,Bor09} (2.6\% of the force between ideal metals,
$\sim\pi\alpha$).

In this paper we neglect the effects due to $\alpha$ on the force. In
this approximation the force between a suspended graphene sheet and
any another material tends to zero at $T= 0$ (negligible in
comparison with the force between bulk materials). If graphene covers
a substrate then the force difference $\Delta F=F_g-F_b$ is equally
negligible, where $F_g$ and $F_b$ are the the force with and without
the graphene layer on the substrate, respectively. One can
systematically neglect the effects $\sim \alpha$ in $\Delta F$ by
taking the non-retarded limit $c\rightarrow\infty$. The possibility
to use this limit was already indicated for two graphene sheets
\cite{Gom09}. Detailed calculation of the force between suspended
graphene and Au \cite{Fia11} gave an independent proof of this
approximation. Taking the limit $c\rightarrow \infty$ in the Lifshitz
formula one finds the graphene contribution:
\begin{equation}\label{DF}
    \Delta F(a,T)=\frac{T}{8\pi
    a^3}{\sum\limits_{n=0}^{\infty}}^{\prime}
    \int\limits_{\xi_n}^{\infty}dxx^2\left[\frac{R }{e^x-R }-
    \frac{R_0}{e^x-R_0}\right],
\end{equation}
where the integration variable in the physical terms is $x=2aq$. Here
$R=r_{1}r_{2}$ is the product of the reflection coefficients for the
body 1 (covered with graphene) and the body 2, and $R_0=r_{0}r_{2}$,
where $r_0$ is the reflection coefficient of the body 1 without
graphene. The reflection coefficients also have to be calculated in
the non-retarded limit. The sum is taken over the imaginary Matsubara
frequencies $\omega_n=2i\pi Tn/\hbar$, which enter the dielectric
functions in the reflection coefficients. Only p polarization
contributes to $\Delta F$ since the s polarization vanishes in the
non-retarded limit. It has to be stressed that $c\rightarrow\infty$
limit can only be applied to the force difference but not to $F_g$ or
$F_b$ separately. We keep the lower integration limit in (\ref{DF})
finite $\xi_n=2\pi Tn (\hbar c/2a)^{-1}$. Doing so we stay within
acceptable uncertainty $\sim \alpha$ in $\Delta F$. This definition
is more convenient because convergence of $\Delta F$ is defined only
by graphene but not high frequency transparency of the bulk bodies.

To proceed further we need to know the dielectric function of
graphene. It is related to the dynamical conductivity of the
vacuum-graphene-dielectric system by the relation \cite{Ste67}:
\begin{equation}\label{eps_g}
    \varepsilon(q,\omega)=1+\frac{4\pi\sigma(q,\omega)}{\omega}
    \left(\frac{k_0k_1}{\varepsilon_1 k_0+k_1}\right).
\end{equation}
Combining Eq. (\ref{eps_g}) with Eq. (\ref{rp}) one finds a simple
expression for the reflection coefficient of the body covered with
graphene:
\begin{equation}\label{r1_eps}
    r_1=1-\frac{1-r_0}{\varepsilon(q,\omega)}.
\end{equation}

\section{Dielectric function of graphene}

The dielectric function of graphene can be calculated using the
random phase approximation (RPA). The RPA was used extensively for
graphene in different situations (see the reviews
\cite{Cas09,Kot10}). Specific to our case, we need to know this
function for imaginary frequencies at nonzero temperature for doped
graphene with a nonzero gap. In the literature one can find
$\varepsilon(q,\omega)$ only in different limiting cases.

For nonzero gap the electron energy in the valence ($s=-1$) or in the
conduction ($s=+1$) band is $E_{s\textbf{k}}=s\sqrt{(\hbar v_F
\textbf{k})^2+\Delta^2}$. The probability to find an electron (hole)
with the energy $E_{s\textbf{k}}$ is given by the Fermi distribution
$f_{s\textbf{k}}=[1+e^{(E_{sk}-E_F)/T}]^{-1}$, where $E_F$ is the
Fermi level. In the RPA, the dielectric function of graphene can be
expressed as $\varepsilon=1+v_c(q)\Pi(q,\omega)$. Here $v_c=2\pi
e^2/\kappa q$ is the 2D Coulomb interaction, $\kappa$ is defined by
the environment of the graphene layer (in our case
$2\kappa=\varepsilon_1(0)+1$), and $\Pi(q,\omega)$ is the 2D
polarizability given by the bare bubble diagram:
\begin{equation}\label{Pi_def}
    \Pi(q,\omega)=-4\sum\limits_{s,s'}\int\frac{d^2k}{(2\pi)^2}
    V^{ss'}_{kk'}
    \frac{f_{s\textbf{k}}-f_{s'\textbf{k}'}}{\hbar\omega+E_{s\textbf{k}}
    -E_{s'\textbf{k}'}},
\end{equation}
where $\textbf{k}'=\textbf{k}+\textbf{q}$, $s,s'=\pm 1$, and the
vertex factor is given by $2V^{ss'}_{kk'}=1+(\hbar^2
v_F^2\textbf{k}\cdot \textbf{k}'+\Delta^2)/E_{sk}E_{s'k'}$. The
factor 4 at the front comes from two spins and two valleys
degeneracy.

In what follows we use the dimensionless variables:
\begin{equation}\label{var_dimless}
    Q=\frac{\hbar v_Fq}{2T},\ \ \ Z=\frac{\hbar\zeta}{2T},\ \ \
    \Delta_T=\frac{\Delta}{T},\ \ \ \epsilon_F=\frac{E_F}{T},
\end{equation}
where $\zeta$ is the imaginary frequency. It is convenient to
calculate the polarizability in the elliptic coordinates $\mu$ and
$\nu$ defined by the relations
\begin{equation}\label{ellip}
    k=\frac{q}{2}\left(\cosh\mu-\cos\nu\right),\ \
    k'=\frac{q}{2}\left(\cosh\mu+\cos\nu\right).
\end{equation}
The notations $\xi=\cosh\mu$ and $\eta=\cos\nu$ will also be used.
Separating interband ($\textbf{k}$ and $\textbf{k}'$ in different
bands) and intraband ($\textbf{k}$ and $\textbf{k}'$ in one band)
transitions in (\ref{Pi_def}) we can present the dielectric function
as:
\begin{equation}\label{eps}
    \varepsilon(q,i\zeta)=1+\frac{\alpha_g}{\pi}\left(I_1+I_2\right),\
    \ \ \alpha_g=\frac{e^2}{\kappa\hbar v_F},
\end{equation}
where $I_1$ and $I_2$ are the contributions coming from interband
and intraband transitions, respectively, and $\alpha_g$ is the
interaction constant in graphene. For $I_{1,2}$ one finds
\begin{widetext}
\begin{equation}
  I_{1,2} = \int\limits_0^{\infty}d\mu\int\limits_0^{\pi}d\nu
  \left[1\mp\frac{Q^2(\xi^2+\eta^2-2)+\Delta_T^2}
  {\epsilon_1\epsilon_2}\right]\frac{Q(\epsilon_2\pm\epsilon_1)
  \left(\xi^2-\eta^2\right)}{4Z^2+(\epsilon_2\pm\epsilon_1)^2}
  \left[\frac{\sinh\epsilon_2}{\cosh\epsilon_2+\cosh\epsilon_F}\pm
  \frac{\sinh\epsilon_1}{\cosh\epsilon_1+\cosh\epsilon_F}\right].
  \label{I1}
\end{equation}
\end{widetext}
\begin{floatequation}
\mbox{\textit{see eq.~\eqref{I1}}}
\end{floatequation}
In eq. (\ref{I1})
$\epsilon_{1,2}=\sqrt{Q^2(\xi\mp\eta)^2+\Delta_T^2}$ and the upper
(lower) sign is related to index 1 (2).

Typical values of $q$ for the Casimir problem are $\sim 1/2a$.
Therefore, for the distances $a\gg\xi_T$ of interest in this paper,
the values of $Q$ are always small, i. e. $Q\ll 1$. In this limit eq.
(\ref{I1}) can be simplified further. The parameter $Q\eta$ is always
small but $Q\xi$ is not. In fact, the important values of $\xi$ in
the integrals are large: $\xi\sim\max(1/Q,\Delta_T/Q)$. Making the
corresponding expansions and performing explicit integrations over
$\nu$ we find for $I_{1,2}$ in the limit $Q\ll 1$:
\begin{equation}\label{I1_small}
    I_1=\pi Q\int\limits_{\Delta_T}^{\infty}d\epsilon
    \frac{\epsilon^2+\Delta_T^2}{\epsilon^2(Z^2+\epsilon^2)}
    \cdot\frac{\sinh\epsilon}{\cosh\epsilon+\cosh\epsilon_F},
\end{equation}
\begin{eqnarray}\label{I2_small}
    \nonumber
    I_2 &=& \frac{2\pi}{Q}\int\limits_{\Delta_T}^{\infty}d\epsilon
    \epsilon\left[1-\frac{Z}{\sqrt{Z^2+Q^2-(\Delta_T Q/\epsilon)^2}}
    \right]\times \\
    &&\frac{1+\cosh\epsilon\cosh\epsilon_F}
    {\left(\cosh\epsilon+\cosh\epsilon_F\right)^2},
\end{eqnarray}
where we introduced a new integration variable
$\epsilon=\sqrt{Q^2\xi^2+\Delta_T^2}$. Note that the intraband
contribution dominates the dielectric function in the $Q\ll 1$
limit.

\section{The force}

In the large distance limit $a\gg\xi_T$ the dielectric function of
graphene is significant ($\varepsilon-1\gg\alpha$) at frequencies
$\hbar\zeta\lesssim T$, which are low for $T\sim 300^{\circ}$ K. For
these frequencies most of dielectric materials have static
permittivities and metals can be considered as perfect conductors. In
such cases we can simplify the calculation of $\Delta F$ in eq.
(\ref{DF}) taking the static permittivities $\varepsilon_{1,2}(0)$
for bulk bodies ($\varepsilon_{2}(0)\rightarrow\infty$ for metals)
and keeping $q$ and $\zeta$ dependence only for the graphene
dielectric function $\varepsilon(q,i\zeta)$. It has to be mentioned
that $\varepsilon(q,i\zeta)$ is essentially nonlocal. This
nonlocality, however, is two dimensional, which simplifies the
calculation of the Casimir force in comparison with the 3D case
\cite{Esq04}. This is because there is only an in-plane wave vector.

\begin{figure}[ptb]
\begin{center}
\includegraphics[width=0.48\textwidth]{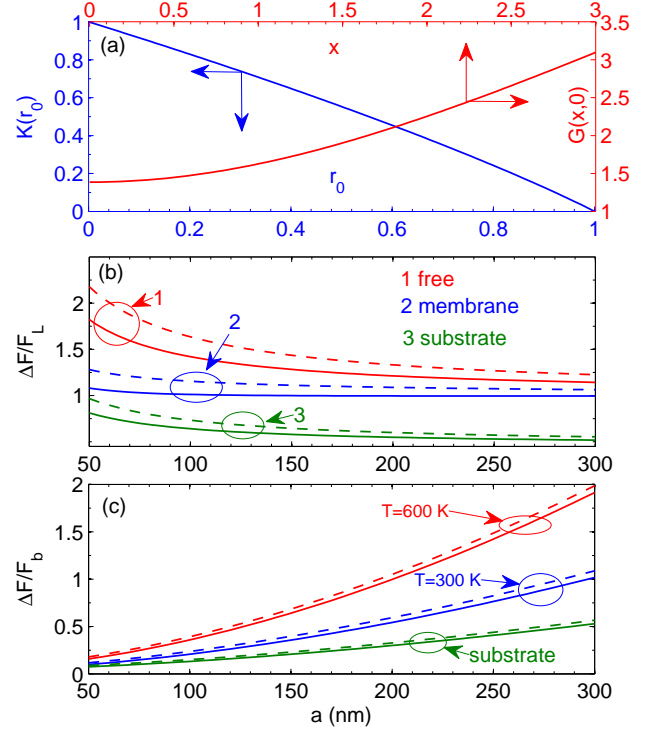}
\vspace{-0.3cm} \caption{(Color online) (a) (top-right axes)
Function $G(x,0)$ that enters Eq. (\ref{DED0}).
(bottom-left axes) The factor $K(r_0)$ in Eq. (\ref{K_def}) as a
function of the static reflection coefficient of the substrate
supporting graphene. (b) The force ratio $\Delta F/F_L$ as a
function of distance for free standing graphene,
graphene-on-membrane, and graphene-on-substrate at $T=300$ K. The
solid lines are for $\epsilon_F=0$ and the dashed lines are for
$\epsilon_F=10$. (c) The relative force for graphene-on-membrane
as a function of $a$ for two different temperatures. The lowest
dashed and solide curves are for graphene-on-substrate at $T=300$ K.
The curves for membrane in (b) and (c) were calculated for $h=20$
nm and $r_0=0.6$.} \label{fig1}
\vspace{-0.5cm}
\end{center}
\end{figure}

Consider first a gapless graphene. For $\Delta=0$ the dielectric
function at large distances $a\gg\xi_T$ follows from (\ref{eps}) and
(\ref{I2_small})
\begin{equation}\label{DED0}
    \varepsilon(q,i\zeta)=1+\frac{2\alpha_gG(\epsilon_F,0)}{Q}
    \left(1-\frac{Z}{\sqrt{Z^2+Q^2}}\right).
\end{equation}
The function $G(\epsilon_F,0)$ here increases monotonously
starting from $2\ln 2$ at $\epsilon_F=0$ (fig. \ref{fig1}a). In
general $G(x,y)$ is given by the expression
\begin{equation}\label{G_def}
    G(x,y)=\int\limits_{y}^{\infty}dtt
    \frac{1+\cosh t\cosh x}
    {\left(\cosh t+\cosh x\right)^2}.
\end{equation}
Let us stress that the characteristic frequency in the dielectric
function (\ref{DED0}) is $\zeta \sim v_Fq$ as is expected from
general consideration \cite{Gom09}. For the Matsubara frequency
$\omega_0=i0$ the dielectric function has a metallic character, i. e.
$\varepsilon(q,i0)\gg 1$ and the reflection coefficient of the body
covered with graphene approaches 1, i. e. $r_1\rightarrow 1$.
Already, for $n=1$ we have $Z_1\gg Q$ and $\varepsilon(q,i\zeta_1)$
is strongly suppressed. For $n\neq 0$ the reflection coefficient
approaches the substrate value, $r_1\rightarrow r_0$. Let us stress
that just one monolayer covering the substrate makes it perfectly
reflecting at low frequencies.

%The physical explanation for this behavior \cite{Gom09} comes from
%the excitation of low frequency plasmons in the graphene sheet.

For distances $a\gg \xi_T$ we can apply (\ref{DED0}) to calculate
the force (\ref{DF}). The $n=0$ term dominates in $\Delta F$. If the
second body is a metal we can take $R=1$ and $R_0=r_{0}$, where
$r_0$ has to be taken in the static limit. The force in this case
is:
\begin{equation}\label{K_def}
    \Delta F(a,T)=\frac{T\zeta(3)}{8\pi a^3}K(r_0), \ \ \ a\gg \frac{\hbar
    v_F}{T},
\end{equation}
where the function $K(r_0)$ describes the effect of the substrate
on the force. This function is shown in fig.\ref{fig1}a and is
expressed analytically as
\begin{equation}\label{K_expr}
    K(r_0)=\frac{1}{2\zeta(3)}\int\limits_0^{\infty}{\textrm d}xx^2
    \left[\frac{1}{e^x-1}-\frac{r_0}{e^x-r_0}\right].
\end{equation}
For suspended graphene $r_0=0$, and Eq. (\ref{K_def}) coincides with
the Lifshitz force $F_L$. Note that a metallic substrate for
graphene will result in the zero force because $K(r_0)\rightarrow 0$
when $r_0\rightarrow 1$.

The effect of graphene will be appreciable if $\Delta F$ is
measurable but also if $\Delta F$ is not negligible in comparison
with the background force $F_b$. The force (\ref{K_def}) is maximal
for free standing graphene when $F_b=0$. This configuration is
realizable in practice \cite{Zan10} and has significant interest.
However, it can not always be practical due to the deformation
induced by the force. A more stable configuration is graphene on a
dielectric membrane of thickness $h$. For a membrane, the reflection
coefficient is
\begin{equation}\label{membr}
    r_{0m}=r_0\frac{1-e^{-2qh}}{1-r_0^2e^{-2qh}},
\end{equation}
where $r_0$ corresponds to the bulk material. For a thin membrane,
$h\ll a$, $r_{0m}$ becomes small and the background force $F_b$ is
much weaker than that for the thick substrate. For
graphene-on-membrane the force in the long distance limit is also
given by eq. (\ref{K_def}) but now the factor $K$ depends slightly
on the distance due to $q$-dependence of $r_{0m}$. The
graphene-on-membrane configuration maximizes not only the absolute
value of the force $\Delta F$ but also the relative value $\Delta
F/F_b$. This is an important practical observation.

%===================================================================================================
%===================================================================================================

Figure \ref{fig1}b shows how the force approaches its limit value
(\ref{K_def}) for free standing graphene, for graphene on 20 nm thick
SiO$_2$ membrane, and for graphene on a thick SiO$_2$ substrate.
Numerical calculations were performed using the dielectric function
(\ref{eps}) with $I_{1,2}$ from (\ref{I1}) without additional
approximations. The continuous lines are for $E_F=0$ and the dashed
lines are for $E_F=10\,T$. One can see that the force is not very
sensitive to $E_F$.

This is especially obvious in fig. \ref{fig1}c where the relative
force ($\Delta F$ in respect to the background force $F_b$) is shown.
This figure demonstrates significant dependence on temperature and
shows that the relative force is considerably smaller for a thick
substrate than for a thin membrane.

Significant dependence on the Fermi level is desirable to change the
force by electronic means. This can be realized if graphene has a
non-zero gap. The material will change from insulating to conducting
state in response to the position of $E_F$. It has to influence the
dielectric function and thus the force. The dielectric function of
graphene with the gap $2\Delta$ was calculated in \cite{Pya09}, on
the real frequency axis at $T=0$. Here we are using our result
(\ref{eps}), (\ref{I1}) for the dielectric function on the imaginary
frequency axis at non-zero $T$.

As in the case of gapless graphene the main contribution to the force
at large distances comes from the $n=0$ term, which depends on the
static dielectric function:
\begin{equation}\label{eps_stat}
    \varepsilon(q,i0)=1+\frac{2\alpha_g}{Q}G(\epsilon_F,\Delta_T),
\end{equation}
where the function $G(\epsilon_F,\Delta_T)$ is given by eq.
(\ref{G_def}). The gap gives significant effect for $\Delta_T\gg 1$.
If the Fermi level is in the middle of the gap, i. e. $\epsilon_F=0$,
the function $G$ is exponentially suppressed, i. e.
$G(0,\Delta_T)\approx 2\Delta_T e^{-\Delta_T}$. In this case the
effect of graphene on the force is small. When $\epsilon_F$ becomes
comparable with $\Delta_T$ the dielectric function
$\varepsilon(q,i0)$ is large and the effect of graphene is
significant. In the long distance limit the force behavior is similar
to eq. (\ref{K_def}).

%===================================================================================================
\begin{figure}[ptb]
\begin{center}
\includegraphics[width=0.48\textwidth]{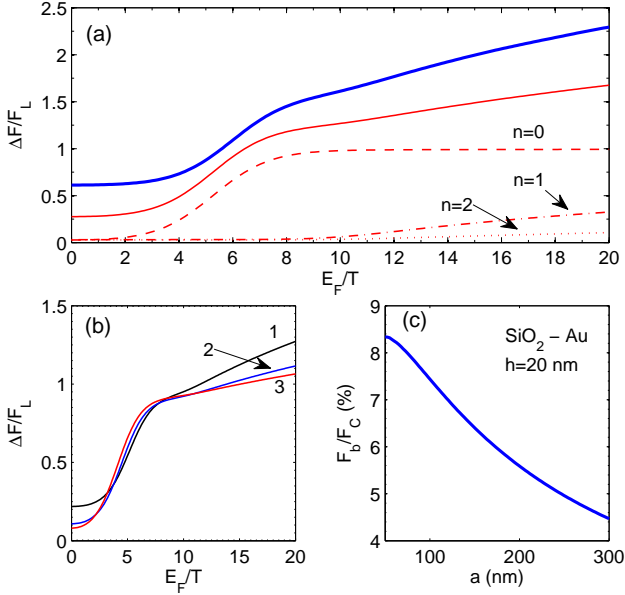}
\vspace{-0.3cm} \caption{(Color online) (a) The force as a
function of the Fermi level for free standing graphene with
$\Delta_T=10$. The thick line is for $a=50$ nm and the thin
solid line is for $a=100$ nm. The dashed, dash dotted, and
dotted lines are the first three components for the $a=100$ nm
case. (b) The force for graphene-on-membrane ($\Delta_T=10$).
The lines marked as 1, 2, and 3 correspond to $a=100$, 200,
and 300 nm, respectively. (c) The background force as a function
of the distance in the units of the bare Casimir force
$F_C=\pi^2\hbar c/240 a^4$.}
\label{fig2} \vspace{-0.5cm}
\end{center}
\end{figure}
%===================================================================================================

Figure \ref{fig2}a shows the force for suspended graphene with the
gap $\Delta_T=10$ as a function of the Fermi level for $a=50$ nm and
100 nm (solid curves). About ten terms are important in the sum
(\ref{DF}); the first three terms for $a=100$ nm are shown. Indeed,
the $n=0$ term gives the main contribution. The finite value of the
force at $E_F=0$ decreases as $a$ and $\Delta$ increase. It is mainly
due to interband transitions, which are not included in
(\ref{eps_stat}). As expected, the force is small for $E_F=0$ and is
on the level of $F_L$ for the Fermi level $E_F\gtrsim \Delta$.
Typically the force changes $3 - 5$ times on the interval
$0<E_F\lesssim \Delta$ proving significant sensitivity to the Fermi
level position.

The force for graphene-on-membrane is shown in fig. \ref{fig2}b. The
behavior is similar to that for suspended graphene. However, in this
case the force has to be compared with the background force for
membrane shown in fig. \ref{fig2}c. The latter one was calculated
using frequency dependent dielectric functions of SiO$_2$ and Au. The
relative force $\Delta F/F_b$ varies in the range $10 -100$\%; it is
small for short separation and increases with $a$. The background
force $F_b$ can be reduced further by decreasing thickness and/or the
permittivity of the membrane.

\section{Conclusions and discussion}

In this paper we analyzed the Casimir interaction of a
graphene-covered dielectric with a metal plate. The dielectric
function of graphene was found at finite temperature for imaginary
frequencies for the material with a finite bandgap and non-zero Fermi
level. A simple expression (\ref{DF}) describes the graphene
contribution to the force. We can conclude that for graphene with the
gap $2\Delta\gg T$ there is a strong dependence of the Casimir force
on both the temperature and the Fermi level. This is realized at
distances $a\gg \hbar v_F/T$ when the main contribution to $\Delta F$
originates from thermal fluctuations. The predicted force is
measurable with modern AFM instruments and can have significant
technological applications. Graphene-on-membrane interacting with a
metal has special interest for practical applications. This
configuration combines mechanical strength with unique electronic
properties of graphene. It allows tailoring of the Casimir force by
electronic means. Manipulations with the thermal force opens up
completely new possibilities which, so far, seemed to have pure
academic interest for condensed matter. For example, it becomes
possible to observe the nonequilibrium Casimir force
\cite{Ant06,Ant08} between solid bodies at distances $\sim 100$ nm.
This possibility put the Casimir effect on the same ground as the
short distance radiative heat transfer \cite{Rou09}. For all bulk
materials the equilibrium component of the force at $a\sim 100$ nm is
orders of magnitude larger than the nonequilibrium one. However, for
suspended graphene or graphene on membrane interacting with a metal
these components of the total force can be comparable.

 %%%%%%%%%%%%%%%%%%%%%%%%%%%%%%%%%%%%%%%%%%%%%%%%%%%%%%%%%%%%%%%%%%%%%%%%%%%%%%%%%%%%%%%%%%%%%%%%

\end{document}